\documentclass[12pt]{article}

\usepackage{graphicx}
\usepackage{epsfig}
\begin{document}
%******************************************************************
\baselineskip=.33in
%******************************************************************

%***********************
\newcommand{\be}{\begin{equation}}
\newcommand{\ee}{\end{equation}}
\newcommand{\bea}{\begin{eqnarray}}
\newcommand{\eea}{\end{eqnarray}}
\newcommand{\da}{\dagger}
\newcommand{\dg}[1]{\mbox{${#1}^{\dagger}$}}
\newcommand{\hlf}{\mbox{$1\over2$}}
\newcommand{\lfrac}[2]{\mbox{${#1}\over{#2}$}}
\newcommand{\scsz}[1]{\mbox{\scriptsize ${#1}$}}
\newcommand{\tsz}[1]{\mbox{\tiny ${#1}$}}
\newcommand{\doref}{\bf (*** REF.??? ***)}

%***************

\begin{flushright} 
% ArXiv \\
LA-UR-07-7264 \\
\end{flushright} 

%******************

\begin{flushleft}

\Large{\bf New Horizons and the Onset of the Pioneer Anomaly}

\vspace{0.5in}

\normalsize
\bigskip 

{\bf Michael Martin Nieto}  \\

\normalsize
\vskip 15pt

Theoretical Division (MS-B285), Los Alamos National Laboratory,\\
Los Alamos, New Mexico 87545, U.S.A. \\
Email: mmn@lanl.gov \\

%**************************************************

\vspace{0.5in}
\bigskip

\end{flushleft}

%********************************************************************
\baselineskip=.33in
%******************************************************************

\begin{abstract}

Analysis of the radio tracking data from the Pioneer 10/11 spacecraft
at distances between about 20 - 70 AU from the Sun has 
indicated the presence of an unmodeled, small, constant, Doppler blue 
shift which can be interpreted as a constant acceleration
of $a_P= (8.74 \pm 1.33) \times 10^{-8}$  cm/s$^2$   
directed approximately {\it towards} the Sun.  In addition, there is early (roughly modeled) data from as close in as 5 AU which indicates there may have been an onset of the anomaly  near Saturn.   We observe that the data now arriving from the New Horizons mission to Pluto and the Kuiper Belt could allow a relatively easy, direct experimental test of whether this onset is associated with distance from the Sun (being, for example, an effect of drag on dark matter).  We strongly urge that this test be done.   \\

\end{abstract}

\begin{center}
\today
\end{center}

\newpage

%********************************** 1

% \section{The Pioneer missions}

\begin{figure}[h!] 
    \noindent
    \begin{center}  
            \epsfig{file=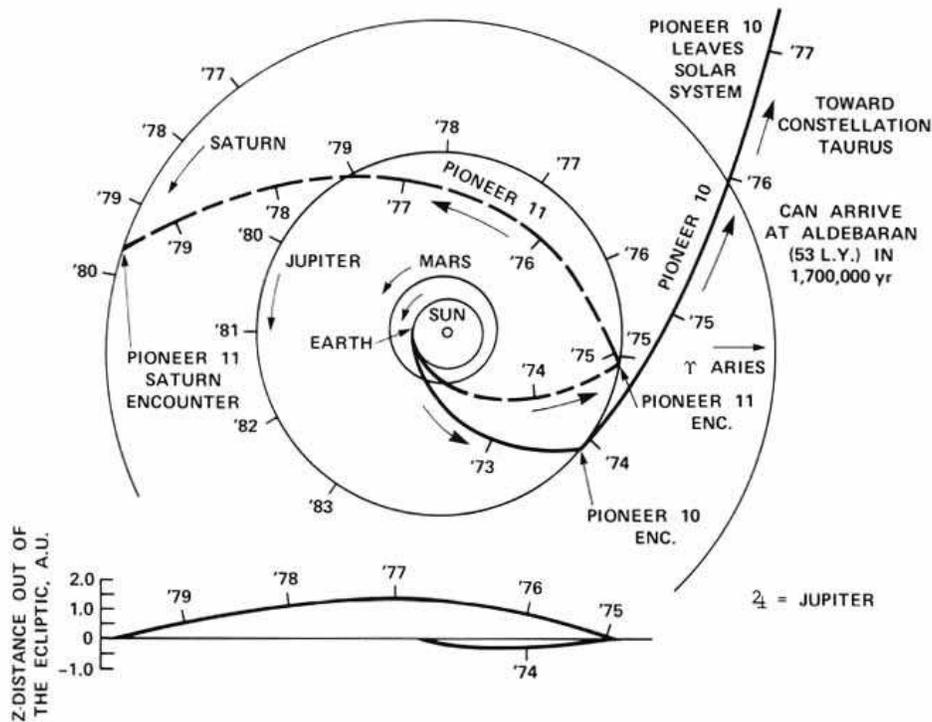,width=5in}%,angle=270}           
\caption{The Pioneer orbits in the interior of the solar system.}
 \label{earlyorbits}
    \end{center}
\end{figure}

Pioneer 10, launched on 3 March 1972 ET (2 March local time),
was the first craft launched into deep space and the first to reach an outer giant planet, Jupiter, on 4 Dec. 1973. 
During its Earth-Jupiter cruise Pioneer 10 was still bound to the
solar system.  With Jupiter encounter, Pioneer 10 reached escape velocity from the solar system, and is now headed in the general direction opposite the relative motion of the solar system in the local interstellar dust cloud.
The Pioneer 10 solar-system orbit is shown in Figure \ref{earlyorbits}.

%******************************** Fig 1

%*****************************************

Pioneer 11 followed soon after with a launch on 6 April 1973 (ET), cruising 
to Jupiter on an heliocentric ellipse. 
On 2 Dec. 1974 Pioneer 11 reached Jupiter, where it underwent the Jupiter gravity assist that sent it back inside the solar system to catch up with Saturn on the far side.   It was then still on an ellipse, but a more energetic one.  

Pioneer 11 reached Saturn on 1 Sept. 1979.  After encounter Pioneer 11 was on an escape hyperbolic orbit.
(Pioneer 11 reached a state of positive solar-system total energy about 2-1/2 hours before closest approach to Saturn.)

The motion of Pioneer 11 is approximately in the direction of the Sun's relative motion in the local interstellar dust cloud (towards the heliopause).  It is roughly anti-parallel to the direction of Pioneer 10.  
Figure \ref{earlyorbits} shows the Pioneer 
11 interior solar system orbit.

%**************************************** 

\begin{figure}[h]
 \begin{center}
\noindent    
\psfig{figure=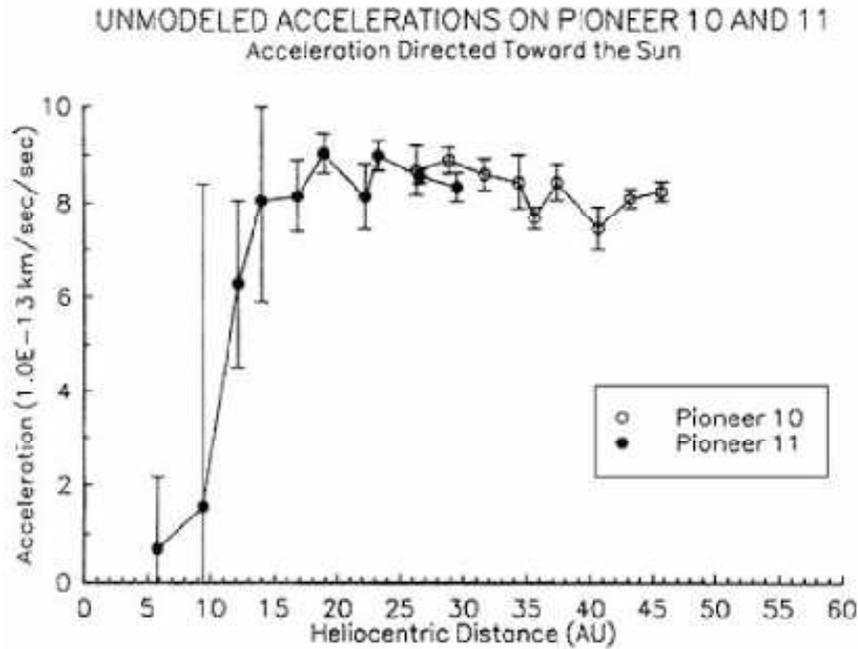,width=115mm}%,height=110mm}
\end{center}
\vskip -10pt
  \caption{A JPL Orbital Data Program (ODP) plot of the early unmodeled 
accelerations of Pioneer 10 and Pioneer 11, from about 
1981 to 1989 and 1977 to 1989, respectively.
\label{fig:correlation}}
\end{figure}

Among the problems encountered in precisely navigating in the interior of the solar system were the intense solar radiation pressure and modeling of the many gas-jet maneuvers.  Even so, with measurements, calibrations, and models, both Pioneers were successfully navigated \cite{null76}.  
After 1976 samples of data were periodically analyzed, to set limits on any unmodeled forces.  (This was especially true for Pioneer 11 which was then on its Jupiter-Saturn cruise.)  At first nothing was found.    
But when a similar analysis was done around Pioneer 11 's Saturn flyby, things dramatically changed.  (See the first two data points in Fig. 
\ref{fig:correlation}.)  So people kept following Pioneer 11.  They also started looking more closely at the incoming Pioneer 10 data.

%************  Fig 2

%**************

The raw data samples for these data points were taken by different individual analysts, averaged over 6 month to 1 year bins, and  the navigational data extracted. Each individual used their own data-editing strategy, models, etc., and the points in Fig. \ref{fig:correlation} were generated from these results.\footnote{The team  included J. D. Anderson, J. Ellis,  E. L. Lau, N. A. Mottinger, G. W. Null, and S. K. Wong.
}
Further, the navigational data was not carefully archived.  That was not really necessary then and, in any event, at first this anomaly was generally believed to only be a ``curiosity." 

By 1992 an interesting string of data-points had been obtained.  They were gathered in a JPL memorandum \cite{JPLmemo}, where Fig. ref{fig:corrrelation} first appeared.  This showed a consistent bias from around 10 AU out corresponding to an acceleration of 
$\sim 8 \times 10^{-8}$ cm/s$^2$.  See Figure \ref{fig:correlation}.

This eventually led to announcement of the effect \cite{bled}, and to a detailed analysis of the data received from 1987.0 to 1998.5 \cite{pioprl,pioprd}.  
The final report \cite{pioprd} extensively addressed
possible sources for a systematic origin for the detected anomaly.
The conclusion was that, even after all {\it known} systematics are
accounted for \cite{pioprd,piompla}, 
at distances between about 20 to 70 AU from the Sun 
there remains an unmodeled frequency drift of 
size $(5.99 \pm 0.01) \times 10^{-9}$ Hz/s.
This drift can be interpreted as an anomalous acceleration signal of
\be 
a_P=(8.74 \pm 1.33) \times 10^{-8} ~~ \mathrm{cm/s}^2, 
\ee
approximately in the direction towards the Sun.  This effect is what has become known as the ``Pioneer anomaly."

% ************* 

Although it is most likely that the anomaly is due to uncalibrated heat systematics, this is not certain.  In particular, suppose 
that the values of the close-in Pioneer 11 data
points in Figure \ref{fig:correlation}
represent a correct rough measurement, and not a problem with signal
to noise.  Then, the huge error at the second data point of
Figure \ref{fig:correlation} evaluated near Saturn flyby  could
represent an onset of the anomaly near Saturn flyby. 
The second Pioneer 11 data point  (whose stated distance of 9.39 AU was on day 1979/244, after 
Saturn encounter) comes from a data span that started before Saturn 
encounter.\footnote{  
The second Pioneer 11 data point was stated to have been taken before (or at) Saturn encounter at 9.39 AU \cite{JPLmemo,edata}.  But since Saturn encounter was at 9.38 AU, that would mean there either was a round-off in the distance quoted or the data overlapped the encounter.  Either way the huge error in this point is anomalous and therefore it is of great interest to reanalyze this region.
} 

One can speculate on a number of reasons why the onset might occur at this distance.  Here, for definiteness, we focus on one, the possibility that there is a cloud of dark matter about the Sun originating near that distance.
Then the anomaly and its onset could represent a drag force \cite{drag} from dark matter of the form \cite{foot}
\be
a_{\tt s}(r)= -\mathcal{K}_{\tt s}~\frac{\rho(r)~v_{\tt s}^2(r)\,A_{\tt s}}{m_{\tt s}}, 
\label{dac}
\ee
where $\rho(r)$ is the density of the (dark) interplanetary medium, 
$\mathcal{K}_{\tt s}$ is the effective reflection/absorption/transmission coefficient of the spacecraft for the 
particles hitting it, $v_{\tt s}(r)$ is the effective relative velocity of the craft with respect to the medium, $A_{\tt s}$ is the effective cross-sectional area of the craft, and $m_{\tt s}$ is its mass.  

In general $\mathcal{K}_{\tt s}$ is between 0 and 2. 
(We take $\mathcal{K}_{\tt s}$ to be a unit constant and the drag 
velocity to be $v_{\tt s} \sim 12$ km/s, about the radial velocity of the 
Pioneers.\footnote{
The precise hyperbolic velocities of Pioneer 10 and 11 are about 12.2 and 11.6 km/s, respectively.
}   
We can consider the effective area to be that of the Pioneers' 
antennae (radii of 1.37 m) and the mass (with half the 
fuel gone) to be 241 kg \cite{pioprd}.  

Given this, the interesting unknown is $\rho(r)$.  An 
axially-symmetric distribution with a {\it constant}, uniform density
that could produce the anomaly would be
\be
\rho_{\tt P}(r)  = 3 \times 
     10^{-19}~\mathrm{g/cm}^3.
\label{rhoP}
\ee
What we know about ordinary dust and gas 
indicates that by far there is not enough of it to yield a sufficient drag force in the well-studied region beyond 20 AU \cite{drag}. 

The point of this note is that if: a) the anomaly is caused by a non-systematic effect, b) the postulated onset near 9.39 AU \cite{edata} is correct, and c) the origin of the onset is a function of the distance from the Sun,
then the current New Horizons mission to Pluto \cite{newH} should be able to verify this rather easily.\footnote{This is in contrast to the possibility that the onset was correlated with the transition to hyperbolic orbit with Saturn encounter at 9.38 AU.
}  

On 19 Jan 2006 the New Horizons mission to Pluto and the Kuiper Belt was launched from Cape Canaveral.  Of relatively low mass ($\sim$478 kg including hydrazine thruster fuel) and with high velocity ($\sim$20-25 km/s), this craft was not designed for precision tracking.  The systematics, especially the heat and gas leak effects. can be large.  This makes measuring the absolute value of any anomaly difficult, although it would be important. 

However to measure a differential effect, like an onset near 9.39 AU, should be relatively easy. 
In the first place, its gravity assist was at Jupiter on 28 Feb 2007.  the last course correction was on 27 Sept. 2007, and another one will not be needed for at least 3 years. Further, the craft will be in spin-stabilization mode for much of the period until soon before the Pluto encounter on  14 July 2015.
Finally, New Horizons will reach the orbit of Saturn in mid-2008.  

Although there will be large heat systematics, they will be approximately the same before and after reaching 9.39 AU.  The falling off of the inverse-square solar radiation pressure is easily separated from any constant residuals.  Finally, any precession maneuvers will be few in number and small, and hence easily modeled.  

Therefore, the Doppler and range data from the period Oct. 2007 to soon after the end of 2008 periods could supply a clear test in the residuals, at some level, of an onset of a Pioneer-like anomaly near 9.39 AU.  (We emphasize that a negative result would not rule out all onset mechanisms, but would some.)   With the factor of 2 increase in mass and also of velocity, here a drag-induced anomaly onset would be approximately a factor of 2 larger (ignoring a roughly similar cross sectional area for the craft).

In summary, the analysis described here would yield an important result in the study of the Pioneer anomaly and it is greatly encouraged.

For a stimulating discussion on this topic I thank V. Alan Kosteleck\'y and I also acknowledge support by the U.S. DOE.

%**************************************

%***************************

\end{document}